\documentclass[aps,twocolumn,superscriptaddress,groupedaddress,nofootinbib]{revtex4-1}
\usepackage{graphicx}
\usepackage{dcolumn}   
\usepackage{bm}       
\usepackage{amssymb}
\usepackage[toc,page]{appendix}
\usepackage{braket}
\newcommand{\del}{\partial}

\usepackage{amsmath}

\begin{document}

\title{Time travel in vacuum spacetimes}

\author{Sandipan Sengupta}
\email{sandipan@phy.iitkgp.ernet.in}
\affiliation{Department of Physics and Centre for Theoretical Studies, Indian Institute of Technology Kharagpur, Kharagpur-721302, INDIA}

\begin{abstract}

The possibility of time travel through the geodesics of vacuum solutions in first order gravity is explored. We present explicit examples of such geometries, which contain degenerate as well as nondegenerate tetrad fields that are sewn together continuously over different regions of the spacetime. 
These classical solutions to the field equations satisfy the energy conditions.
% and their geodesics are associated with finite tidal acceleration everywhere.
%indicates that one may have to revisit the status of causality as a fundamental principle within classical gravity.

%We present non-Einsteinian solutions in first order gravity, whose geodesics allow the possibility of a time travel. These spacetimes, which satisfy the vacuum field equations, contain degenerate as well as nondegenerate tetrad fields at different regions that are sewn together continuously. These classical solutions do not violate the energy conditions and also do not lead to any divergence in the associated curvature two-form fields.

\end{abstract}%%%%%%%%%%%%%%%%%%%%%%%%%%%%%%%%%%%%%%%%%%%%%%%%%%%%%%%%%%%

\maketitle
%%%%%%%%%%%%%%%%%%%%%%%%%%%%%%%%%%%%%%%%%%%%%%%%%%%%%%%%%%%%%%%%%%%%%%%

\section{Introduction}

The quest for spacetimes which could act as `time machine's has captivated the physicists' imagination for a long time. %Even a mere theoretical realization of any such geometry would force us to revisit the very premise where causality is generally believed to be a fundamental principle, irrespective of whether the framework in question is classical or quantum. 
In fact, it is well-known that the Einstein's equations (with or without matter) do admit spacetime solutions that exhibit closed timelike curves, implying a possible realization of time travel (see \cite{godel,misner,tipler,van,morris,gott,ori,krasnikov} and the references within). While some of them violate the energy conditions, others do not. However, the issue whether all among the latter class correspond to matter fields that are known to exist or are stable against quantum fluctuations remain open. This was more or less the spirit, in which a statement about the absence of closed timelike curves in nature has been invoked through the chronology protection conjecture \cite{hawking}. Subsequently, a number of counterexamples have also been discussed in the literature (see \cite{ori,krasnikov} and references therein).

Here we construct spacetime solutions of gravity theory in vacuum, with the special property that their geodesics allow the possibility of a time travel. All these configurations satisfy the first order equations of motion everywhere, which are obtained directly from the variation of Hilbert-Palatini Lagrangian (and can admit solutions with degenerate tetrad \cite{tseytlin,kaul,kaul1}): 
\begin{eqnarray}\label{eom1}
e_{[\mu}^{[K} D_{\nu}(\omega)e_{\alpha]}^{L]}=0,
e_{[\nu}^{[J} R_{\alpha\beta]}^{~KL]}(\omega)=0~.
\end{eqnarray}
These (non-Einsteinian) geometries exhibit both the possible phases of the tetrad fields associated with zero and non-zero determinant, respectively, over different regions of the same spacetime.
% of these spacetimes
%, the tetrad fields exhibitdemonstrate that by including the  phase with degenerate tetrads ($\det e_\mu^I= 0$) in classical gravity \cite{tseytlin,henneaux}, it is possible to  
%As a result, these geometries are essentially non-Einsteinian. 
%Along their geodesics, a trip back in time  is realized through a nonmonotonic behaviour of the proper time. 
%This feature has no analogue in the Einsteinian time machine spacetimes (e.g. the Misner spacetime). 

Spacetimes where the degenerate and nondegenerate phases coexist have also been  explored earlier in several other contexts \cite{bengtsson,horowitz,bengtsson1,madhavan,kaul2,sengupta}. Among a few recent applications, such a framework has been used to construct (non-Einsteinian) solutions representing spacetime-bridges \cite{sengupta} and extensions of the  exterior Schwarzschild geometry \cite{kaul2}.

While vacuum solutions of first order gravity may in general exhibit nontrivial torsion whose origin is purely geometric  \cite{tseytlin,kaul,kaul1,kaul2,sengupta}, the acausal geometries presented here are all torsionfree by construction. These configurations satisfy all the energy conditions. Neither do these exhibit any divergence in the curvature two-form fields for the full spacetime or in the effective lower dimensional curvature scalars associated with the degenerate phase. Their status or role in the quantum theory would not be discussed in this presentation. The perspective here is purely classical, which is not any less intriguing. Let us now turn to the explicit details of these solutions to the first order field equations.

\section{`Flat' geometry}
In this section, we shall exhibit three different `time machine' spacetimes, which correspond to vanishing field-strength everywhere. 

\subsection{ Solution-I:} 
Let us introduce below a spacetime which is divided into three regions and is described by the following metrics:
\begin{eqnarray}\label{metric1}
 ds^2
 %&\equiv& g_{\rho\sigma}dx^{\rho}dx^{\sigma}
 &=&- f^2(x) dt^2 +f'^2(x) dx^2+dy^2+dz^2\mathrm{~~at}~|x|>x_0,\nonumber\\
 &=& 0-F^2(x)dx^2+dy^2+dz^2 \mathrm{~~at}~|x|\leq x_0
\end{eqnarray}
where, each of the coordinates $(t,x,y,z)$ spans the whole real line $(-\infty,\infty)$ and the functions $f(x)~\&~F(x)$ satisfy:
\begin{eqnarray}\label{f(x)}
f(\pm x_0)=0=f'(\pm x_0),~F(x_0)=0.
%~f(u)\rightarrow \infty \mathrm{~as~}v\rightarrow \infty,
\end{eqnarray}
The internal metric used to raise or lower the $SO(3,1)$ indices is given by $\eta_{IJ}=diag[-1,1,1,1]$  everywhere.
At the two regions $|x|>x_0$, the spacetime metric is invertible and becomes equivalent to flat spacetime in Rindler coordinates upon a reparametrization $x\rightarrow u=f(x)$.  The only nonvanishing component of the torsionless spin connection $\omega_\alpha^{~IJ}(e)={\frac{1}{ 2}} \left[ e^{\beta I} 
\del^{}_{[\alpha}e_{\beta]}^J-e^{\beta J}\del^{}_{[\alpha}e_{\beta]}^I -  
e_{\alpha L} e^{\beta I}e^{\sigma J}\del^{}_{[\beta}e_{\sigma]}^L \right]$ are:
\begin{eqnarray}\label{omega1}
\omega_t^{~01}=1,
\end{eqnarray}
while the curvature two-form components $R_{\mu\nu}^{~IJ}$ are all zero. 
Within the patch $|x|\leq x_0$, the tetrad has one null eigenvalue and is not invertible. The  spin-connection fields are chosen to be:
\begin{eqnarray}\label{omega2}
\hat{\omega}_\mu^{~IJ}=0,
\end{eqnarray}
leading to trivial $\hat{R}_{\mu\nu}^{~IJ}$. 
As is straightforward to verify, both the configurations $(e_\mu^I,\omega_\mu^{~IJ})$ and $(\hat{e}_\mu^I,\hat{\omega}_\mu^{~IJ})$ are solutions to the first order equations of motion (\ref{eom1}). This fact, along with the continuity of the basic gauge-covariant fields (i.e. tetrad, torsion and field-strength) at the phase boundaries $x=\pm x_0$ imply that the whole spacetime constructed above is a solution as a whole.

Note that the spin-connection fields as given above are not continuous across $x=x_0$. We do not demand them to be so either, since they are not gauge-covariant fields. However, it is possible to make the connection fields continuous by a $SO(3,1)$ gauge transformation on the fields at $x\leq x_0$:
\begin{eqnarray*}
\hat{e}^{'I}_\mu &=&\Lambda^I_{~L}\hat{e}^{L}_\mu,\\
\hat{\omega}_\mu^{~'IJ}&=& \Lambda^I_{~K}\hat{\omega}_\mu^{~KL}\left(\Lambda^{-1}\right)^{~J}_{L}+ \Lambda^I_{~K}\partial_\mu \left(\Lambda^{-1}\right)^{KJ}.
\end{eqnarray*}
 The appropriate transformation turns out to be a boost:
\begin{eqnarray*}
\Lambda^I_{~J}=\left(\begin{array}{cccc}
\cosh t & \sinh t & 0 & 0\\
\sinh t & \cosh t & 0 & 0\\
0 & 0 & 1 & 0\\
 0 & 0 & 0 & 1  \end{array}\right).
\end{eqnarray*}
The resulting fields read:
\begin{eqnarray*}
\hat{e}^{'0}_x=F(x)\cosh t ,~\hat{e}^{'1}_x=i F(x)\sinh t,~\hat{\omega}_t^{~'01}=1,
\end{eqnarray*}
 while all the remaining field components remain unaffected. Thus, in this gauge, all the components of tetrad and spin-connection fields are manifestly continuous across $x=x_0$.

Let us now evaluate the (torsionless) affine connection components, which are insensitive to the internal gauge rotations and are relevant for the analysis of geodesics. For the invertible and noninvertible metric phases, these are defined respectively as: \begin{eqnarray}\label{gamma}
\Gamma_{\alpha\beta\rho}&=&g_{\rho\sigma}\Gamma_{\alpha\beta}^{~~\sigma}=\frac{1}{2}\left[\del_{\alpha} g_{\beta\rho}+  \del_{\beta} g_{\alpha\rho}-  \del_{\rho}g_{\alpha\beta}\right],\nonumber\\
\hat{\Gamma}_{\alpha\beta\rho}&=&\hat{g}_{\rho\sigma}\hat{\Gamma}_{\alpha\beta}^{~~\sigma}=\frac{1}{2}\left[\del_{\alpha} \hat{g}_{\beta\rho}+  \del_{\beta} \hat{g}_{\alpha\rho}-  \del_{\rho}\hat{g}_{\alpha\beta}\right]
\end{eqnarray}
From these two equations, the components $\Gamma_{\alpha\beta}^{~~\sigma}$ and $\hat{\Gamma}_{\alpha\beta}^{~~\sigma}$ may be determined, respectively.
$\Gamma_{\alpha\beta}^{~~\sigma}$, corresponding to the invertible phase, reduces to the Christoffel connection determined completely by the metric and its inverse. That is not the case for $\hat{\Gamma}_{\alpha\beta}^{~~\sigma}$, though.
The noninvertibility of the metric with $\hat{g}_{t\mu}=0$ gets reflected through the indeterminacy of the affine connection components $\hat{\Gamma}^{~~t}_{\alpha\beta}$.
% which is equivalent to $\hat{\Gamma}_{\alpha\beta t}=0$. 
This essentially implies that within the degenerate region, the null coordinate (t) becomes nondynamical and any evolution along that is equivalent to a gauge (trivial) motion.
The explicit expressions for the nonvanishing affine connection components are given below:
\begin{eqnarray}
&&\Gamma_{ttx}=f(x)f'(x)=-\Gamma_{txt}=-\Gamma_{xtt},~\Gamma_{xxx}=f'(x)f''(x)\nonumber\\
&& \hat{\Gamma}_{xxx}=-F(x)F'(x).
\end{eqnarray}

The geodesic equations at the region $x>x_0$ are defined as:
\begin{eqnarray} \label{geode}
u^\alpha {\cal{D}}_\alpha u^\beta:=u^\alpha\del_\alpha u^\beta+\Gamma^{~~\beta}_{\alpha\rho} u^\alpha u^\rho=0
\end{eqnarray}
where $u^\alpha=\frac{dx^\alpha}{d\lambda}$ is the tangent vector along an affinely parametrized curve $x^\alpha(\lambda)$~. 
%and $\Gamma^{~~\rho}_{\alpha\beta}$ are 
%the affine connection, which reduce to the Christoffel symbol in this case of vanishing torsion: 
%\begin{eqnarray}\label{gamma1}
%\Gamma_{\alpha\beta}^{~~\sigma}=\frac{1}{2}g^{\rho\sigma}\left[\del_{\alpha} g_{\beta\rho}+  \del_{\beta} g_{\alpha\rho}-  \del_{\rho}g_{\alpha\beta}\right]
%\end{eqnarray}
Using the expressions for the metric and affine connections, the equations (\ref{geode}) become:
\begin{eqnarray}
&&f^2(x) \dot{t}^2-f'^2(x)\dot{x}^2-\dot{y}^2-\dot{z}^2=k,\nonumber\\&&f^2(x) \dot{t}=E,~\dot{y}=k_y,~\dot{z}=k_z
\end{eqnarray}
where $E,k_y,k_z$ are constants of motion and $k=1,0,-1$ characterise the timelike, null and spacelike geodesics, respectively. Here we shall be concerned with the null and timelike cases only, for which the solutions read:
\begin{eqnarray}
&&\lambda 
%&=&\pm \frac{f^2(u)}{2E}+const.~ ~~(k=0),\nonumber\\
= \pm \frac{1}{(k+k_y^2+k_z^2)}\left[E^2-(k+k_y^2+k_z^2)f^2(x)\right]^\frac{1}{2}+const.\nonumber\\
&&~
\end{eqnarray}

Within the degenerate region $x\leq x_0$, the physical motion is confined in the $(x,y,z)$ hyperplane, since the geodesic equation along the null coordinate $t$ becomes redundant. The equations for the physical degrees of freedom read:
\begin{eqnarray}
F^2(x)\dot{x}^2-\dot{y}^2-\dot{z}^2=k,~\dot{y}=k_y,~\dot{z}=k_z
\end{eqnarray}
The solution to these equations are given by:
\begin{eqnarray}
\lambda=\pm \frac{1}{(k+k_y^2+k_z^2)^\frac{1}{2}}\int dx~F(x) +const.
\end{eqnarray}

Let us now adopt a specific ansatz for the functions $f(x)$ and $F(x)$, which may be chosen arbitrarily upto the boundary conditions (\ref{f(x)}):
\begin{eqnarray*}
f(x)=\left(\frac{x^2}{x_0^2}-1\right)^n,~F(x)=\sin\left(\frac{\pi x}{x_0}\right)
\end{eqnarray*}
where $n\geq 2$ is an integer. The resulting metric is $C^N$ across $|x|=x_0$, where the integer $N$ is greater than or equal to unity. With the above, the (globally defined) affine parameter along a geodesic associated with the full spacetime is given by:
\begin{eqnarray}
\lambda 
&=& \pm\frac{1}{(k+k_y^2+k_z^2)}\left[E^2-(k+k_y^2+k_z^2)\left(\frac{x^2}{x_0^2}-1\right)^{2n}\right]^\frac{1}{2}\nonumber\\
&& ~~(\mathrm{at~~}|x|>x_0),\nonumber\\
&=& \pm\frac{x_0}{\pi(k+k_y^2+k_z^2)^\frac{1}{2}}\cos\left(\frac{\pi x}{x_0}\right)~~
(\mathrm{at~~}|x|\leq x_0),
\end{eqnarray}
where the integration constant, which refers to the origin of the affine parameter, has been chosen to be zero.
The demand that the affine parameter must be continuous across $x=\pm x_0$ fixes the constant $x_0$ (size of the degenerate core) as $x_0=\frac{\pi E}{(k+k_y^2+k_z^2)^\frac{1}{2}} $. Clearly, the affine parameter at the core has turning points at $x_*=-Nx_0$ defined by $F(x_*)=0$, where $N$ is a positive integer.
%, and also at $u=\pm u_0$. 
Hence, a massive particle, through a trip within the degenerate core, can come back to the same value of proper time it had started at.
%, which has a global meaning for these solutions. 
The affine parameter for a photon exhibits a similar feature, although it is not clear whether that really implies a time travel or not. 

In general, these geodesics are not closed. In the special case with $y=const.=z$ ($k_y=0=k_z$), a timelike trajectory between two adjacent turning points may be interpreted as a closed curve in the $(\lambda,y,z)$ `spacetime'. The spacetime manifold is not time-orientable at these (isolated) turning points. Even though the coordinate velocity $\frac{du}{d\lambda}$ apparently diverges there, the norm is finite: $u^\alpha u_\alpha=-1$. Hence, one could in principle work in a coordinate system where such a divergence would be absent. 
%While this particular path corresponds to a unit winding number, a longer trip from $u=\frac{u_0}{2}$ to $u=-\frac{N u_0}{2}$ (where $N$ could be any odd integer)  wraps the $\lambda$-circle $N$ times.

It is important to note that the curvature two-form fields associated with these geometries, which essentially encode the tidal accelerations, are trivial everywhere ($-\infty<x<\infty$). Due to the non-invertibility of the metric at $|x|\leq x_0$, one cannot construct curvature scalars for the four-geometry in this region. However, the fact that neither the tetrad nor the connection fields associated with this degenerate phase depend on the null coordinate $t$ implies that this patch may as well be interpreted as a three-geometry, described completely by the nondegenerate part of the four-metric. All the associated three-curvature scalars are manifestly trivial in this case. Finiteness of the curvature two-form as well as of these lower dimensional curvature scalars may be understood as a reflection of the absence of any curvature singularity in these spacetimes.
%, interpretation is allowed in view of the fact that  All the corresponding three curvature scalars are also trivial in this case.
%In the special case of $y=const.=z$, a trip from $u=u_0$ to $u=-u_0$ can be interpreted as a closed timelike curve, since that represents a closed loop along the $\lambda$ (timelike) direction at the $y=const.=z$ hyperplane. 
 
\subsection{Solution-II} 

Here, we consider a different spacetime in the same set of coordinates introduced earlier, such that the hypersurface $x=x_0$ divides it into two regions with invertible and noninvertible metrics, respectively:
\begin{eqnarray}\label{metric3}
 ds^2
 %&\equiv& g_{\rho\sigma}dx^{\rho}dx^{\sigma}
 &=&- f^2(x) dt^2 +f'^2(x) dx^2+dy^2+dz^2\mathrm{~~at}~x>x_0,\nonumber\\
 &=& 0-F^2(x)dx^2+dy^2+dz^2 \mathrm{~~at}~x\leq x_0
\end{eqnarray}
where, the functions $f(x)$ and $F(x)$ satisfy:
\begin{eqnarray}\label{fF1}
f(x_0)=0=f'(x_0),~F(x_0)=0.
%~f(u)\rightarrow \infty \mathrm{~as~}v\rightarrow \infty,
\end{eqnarray}
Adopting the same ansatz $\hat{\omega}_\mu^{~IJ}=0$ as earlier for the (torsionless) spin connection in the degenerate phase, we obtain a vanishing field strength $\hat{R}_{\mu\nu}^{~IJ}=0$. Continuity of the gauge covariant fields is trivially satisfied across $x=x_0$ and the configuration represents a solution to the first order field equations everywhere.

Next, let us choose the metric functions in this case as:
\begin{eqnarray}
&&f(x)=\left(\frac{x}{x_0}-1\right)^n,\nonumber\\
&&F(x)=-(n+1) \left(\frac{x}{x_0}-1\right)^n e^{\left(\frac{x}{x_0}-1\right)^{n+1}},
\end{eqnarray}
where $n\geq 2$ is an even integer. The corresponding metric is $C^{2n-3}$ across the phase boundary.
Proceeding exactly as in the earlier example, we find the solution for the affine parameter to be:
 \begin{eqnarray}
\lambda 
&=& \pm\frac{1}{(k+k_y^2+k_z^2)}\left[E^2-(k+k_y^2+k_z^2)\left(\frac{x}{x_0}-1\right)^{2n}\right]^\frac{1}{2}\nonumber\\
&& ~~(\mathrm{at~~}x>x_0),\nonumber\\
&=& \pm\frac{x_0}{ (k+k_y^2+k_z^2)^\frac{1}{2}}e^{\left(\frac{x}{x_0}-1\right)^{n+1}}~~
(\mathrm{at~~}x\leq x_0)
\end{eqnarray}
In the above, continuity of the affine parameter across $x=x_0$ leads to the constraint $x_0=\frac{E}{(k+k_y^2+k_z^2)^{\frac{1}{2}}}$. The above expression reveals that the affine parameter  is in fact nonmonotonic, having an extrema at the degenerate interface $x=x_0$. In other words, an observer, provided she can cross the surface $x=x_0$ along a geodesic, can travel back to the past in her proper time.
% It becomes zero both at $x=x_0+x_0^{1+\frac{1}{n}}$ and $x\rightarrow -\infty$, 
 %Hence, both the timelike or null geodesic traversing this region admits the possibility of a time travel.
 % In this case, a timelike geodesic with $y=const.=z$ ($k_y=0=k_z$) and going from $x=x_0+x_0^{1+\frac{1}{n}}$ to $x\rightarrow -\infty$ may be interpreted as a closed loop in the $(\lambda,y,z)$ hyperplane.

\subsection{Solution-III}
We shall now present yet another `time-machine' solution with flat gauge potential, such that the null eigenvalue lies along the $x$-direction. Let us introduce the two metrics associated with the zero and non-zero determinant phases as:
\begin{eqnarray}\label{metric3}
 ds^2
 %&\equiv& g_{\rho\sigma}dx^{\rho}dx^{\sigma}
 &=&-  dt^2 +t^2 dx^2+dy^2+dz^2\mathrm{~~at}~t>0,\nonumber\\
 &=& -F^2(t)dt^2+0+dy^2+dz^2 \mathrm{~~at}~t\leq 0,
\end{eqnarray}
subject to the following boundary conditions:
\begin{eqnarray}\label{fF}
F(0)=1,~\dot{F}(0)=0.
%~f(u)\rightarrow \infty \mathrm{~as~}v\rightarrow \infty,
\end{eqnarray}
The internal metric is given by $\eta_{IJ}=diag[-1,1,1,1]$.
The associated field-strength vanishes everywhere:
\begin{eqnarray*}
R_{\mu\nu}^{~IJ}=0=\hat{R}_{\mu\nu}^{~IJ}
\end{eqnarray*}
For $t>0$, the affine connection components have only the following components that are non-vanishing:
\begin{eqnarray}
\Gamma_{txx}=t=\Gamma_{xtx},~\Gamma_{xxt}=-t.
\end{eqnarray}
With these, the equations for timelike and null geodesics become:
\begin{eqnarray}
\dot{t}^2-t^2 \dot{x}^2-\dot{y}^2-\dot{z}^2=k,~t^2 \dot{x}=k_x,~\dot{y}=k_y,~\dot{z}=k_z
\end{eqnarray}
These are solved as:
\begin{eqnarray}
\lambda=\pm \frac{1}{(k+k_y^2+k_z^2)} [(k+k_y^2+k_z^2)t^2+k_x^2]^{\frac{1}{2}} +const.
\end{eqnarray}
For $t\leq 0$ with a degenerate metric, all the affine connection components are found to be trivial, except:
\begin{eqnarray}
\hat{\Gamma}_{ttt}=-F(t)\dot{F}(t).
\end{eqnarray}
While this leads to no dynamics along the $x$-direction, the physical motion is described by the remaining three coordinates through the respective geodesic equations:
\begin{eqnarray}
F^2(t)\dot{t}^2-\dot{y}^2-\dot{z}^2=k,~\dot{y}=k_y,~\dot{z}=k_z,
\end{eqnarray}
These imply:
\begin{eqnarray}\label{lambda2}
\lambda=\pm \frac{1}{(k+k_y^2+k_z^2)^{\frac{1}{2}}}\int dt~F(t)  +const.
\end{eqnarray}

Next, let us adopt the following choice as a prototype:
\begin{eqnarray}
F(t)=1+t^n e^{-\frac{t^2}{2}}
\end{eqnarray}
where $n\geq 3$ is an odd integer. This results in an expression for the affine parameter (\ref{lambda2}) in terms of the Gamma function.
  Noting that $F(t)$ has an extrema at $t=-\sqrt{n}$ 
%and that $F(0)=F(-\infty)=1$, 
we conclude that the timelike geodesic ($k=1$) admits the possibility of a time travel for $t<0$.

Our discussion of flat potential solutions in vacuum gravity ends here. Note that some of the tetrad fields are imaginary for solutions I and II, whereas they are all real for solution III. However, the physical (SO(3,1) invariant) fields which could be constructed from the basic ones, namely the metric $g_{\mu\nu}$, $\Gamma_{\mu\nu\alpha}$ and $R_{\mu\nu\alpha\beta}=R_{\mu\nu}^{~~IJ} e_{\mu I} e_{\nu J}$, are all real for all three solutions.

\section{`Schwarzschild' geometry }

As our final example, we shall construct a spacetime solution of first-order gravity for which the field strength tensors are not trivial. This is represented by the chart $\left(t\in (-\infty,\infty),~u\in 
(-\infty,\infty),~\theta\in[0,\pi],~\phi\in[0,2\pi]\right)$ as:
\begin{eqnarray}\label{sch1}
ds^2 &=&-\left[1-\frac{2M}{f(u)}\right]dt^2 +
\left[1-\frac{2M}{f(u)}\right]^{-1} f'^{2}(u) du^2 \nonumber\\
&+& 
f^2(u)\left[d\theta^2+\sin^2 \theta d\phi^2\right] \mathrm{~~at}~u>u_0,\nonumber\\
&=& 0- F^2(u) du^2 + H^2(u)\left[d\theta^2+\sin^2 
\theta d\phi^2\right]\nonumber\\
&& ~\mathrm{at}~u\leq u_0,
\end{eqnarray}
%Manifestly, the geometry is $R^2\otimes S^2$ at $u>u_0$ and $R\otimes S^2$ at $u\leq u_0$. 
with the boundary conditions:
\begin{eqnarray}\label{fF2} 
f(u_0)=2M,~f'(u_0)=0,~F(u_0)=0.
\end{eqnarray} 
As earlier, the internal metric is given by $\eta_{IJ}=diag[-1,1,1,1]$.
The metric at $u>u_0$ may be brought to the Schwarzschild form through a reparametrization $u\rightarrow r=f(u)$. However, the metric at $u\leq u_0$ is degenerate and has no semblance to the Schwarzschild interior.
A different construction based on metrics of the form above has been discussed earlier, namely, in the context of possible extension(s) of the Schwarzschild exterior based on torsional degenerate geometries
 \cite{kaul2}.
 
The nonvanishing components of the associated (torsionless) spin-connection 
fields $\omega_\alpha^{~IJ}$ are given by:
\begin{eqnarray}\label{omega}
&&\omega_t^{~01}=\frac{M}{f^2(u)},~\omega_\theta^{~12}=-\left(1-\frac{2M}
{f(u)}\right)^{\frac{1}{2}},\nonumber\\
&&\omega_\phi^{~23}=-\cos\theta,~
\omega_\phi^{~31}=\left(1-\frac{2M}{f(u)}\right)^{\frac{1}{2}}\sin\theta 
\end{eqnarray}
Using these, the field strength tensors $R_{\mu \nu}^{~~IJ}(\omega)$ can be  
evaluated to be:
\begin{eqnarray}\label{R}
R_{tu}^{~01}&=&\frac{2Mf'(u)}{f^3(u)},~R_{t\theta}^{~02}
(\omega)=-\frac{M}{f^2(u)}\left(1-\frac{2M}{f(u)}\right)^{\frac{1}
{2}},\nonumber\\R_{t\phi}^{~03}&=&-\frac{M}
{f^2(u)}\left(1-\frac{2M}{f(u)}\right)^{\frac{1}{2}}
\sin\theta ,\nonumber\\
R_{u\theta}^{~12}&=&-\frac{Mf'(u)}{f^2(u)}\left(1-\frac{2M}
{f(u)}\right)^{-\frac{1}{2}},~R_{\theta\phi}^{~23}
= \frac{2M}{f(u)}\sin\theta,\nonumber\\ 
R_{\phi u}^{~31}&=&-\frac{Mf'(u)}{f^2(u)}\left(1-\frac{2M}{f(u)}
\right)^{-\frac{1}{2}}\sin\theta
\end{eqnarray}

 At the region $u\leq u_0$ with a degenerate phase, the torsionless spin-connection 
 has the following nonvanishing components:
\begin{eqnarray}\label{R3}
\hat{\omega}_\theta^{~12}=i\frac{H'(u)}
{F(u)},~\hat{\omega}_\phi^{~23}=-\cos\theta,~\hat{\omega}_\phi^{~31}
=-i\frac{H'(u)}{F(u)}\sin\theta
\end{eqnarray}
The corresponding $SO(3,1)$ field strength components read: 
\begin{eqnarray}\label{Rhat}
&&\hat{R}_{u\theta}^{~12}=i\left[\frac{H'(u)}
{F(u)}\right]',~\hat{R}_{\theta\phi}^{~23}
=\left[1+\left(\frac{H'(u)}
{F(u)}\right)^2\right]\sin\theta,\nonumber\\
&&\hat{R}_{\phi u}^{~31}=i\left[\frac{H'(u)}{F(u)}\right]'\sin\theta,
\end{eqnarray}
all other components being zero. Note that even though some of the field-strength components are imaginary, the physical fields, given by their $SO(3,1)$ gauge-invariant counterparts $\hat{R}_{\mu\nu\alpha\beta}=\hat{R}_{\mu\nu}^{~~IJ} \hat{e}_{\mu I} \hat{e}_{\nu J}$, are all real.

Since this configuration with degenerate tetrads have vanishing torsion by construction, the first of the set of equations of motion (\ref{eom1}) is already satisfied. The remaining equation involving the curvature two-form is also satisfied provided the fields obey the constraint:
\begin{eqnarray}\label{mc}
\left(1+\frac{H'^2(u)}{F^2(u)}\right)F(u)+2H(u)\left(\frac{H'(u)}{F(u)}\right)'=0
\end{eqnarray}
Next, guessing from what the explicit expressions (eqs. (\ref{R}) and (\ref{Rhat})) for the  curvature two-forms suggest, let us adopt an ansatz that ensures the continuity of all the gauge-covariant fields at $u=u_0$:
\begin{eqnarray}\label{hfu}
\frac{H'(u)}{F(u)}=-\left(\frac{2M}{f(u)}-1\right)^{\frac{1}{2}}
\end{eqnarray}
Eqs.(\ref{mc}) and (\ref{hfu}) can be solved explicitly for the functions $F(u)$ and $H(u)$, leading to:
\begin{eqnarray}
&&F(u)=-\frac{f'(u)}{\left(\frac{2M}{f(u)}-1\right)^{\frac{1}{2}}},\nonumber\\
&&H(u)=f(u)
\end{eqnarray}
As discussed earlier, the apparent discontinuity in the connection field ($\omega_t^{~01}\neq \hat{\omega}_t^{~01}$ at $u=u_0$) does not imply a real pathology, and could be gauged away by a boost of the form:
\begin{eqnarray*}
\Lambda^I_{~J}=\left(\begin{array}{cccc}
\cosh \left(\frac{t}{4M}\right) & \sinh \left(\frac{t}{4M}\right) & 0 & 0\\
\sinh \left(\frac{t}{4M}\right) & \cosh \left(\frac{t}{4M}\right) & 0 & 0\\
0 & 0 & 1 & 0\\
 0 & 0 & 0 & 1  \end{array}\right)
\end{eqnarray*}

Let us now analyse the geodesics of the  spacetime described by the field configuration given above. These curves are defined in terms of the affine connection for the two different phases of the tetrad (defined in eq.(\ref{gamma})), whose nontrivial components are displayed below:
\begin{eqnarray}\label{gamma2}
u>u_0 &:&\nonumber\\
\Gamma_{ttu}&=&\frac{Mf'(u)}{f^2(u)},~\Gamma_{tut}=-\frac{Mf'(u)}
{f^2}=\Gamma_{utt},\nonumber\\
\Gamma_{uuu}&=&\frac{1}{2}\del_u 
\left[\frac{f(u)f^{'2}(u)}{f-2M}\right],\nonumber\\
\Gamma_{\theta\theta u}&=&-f(u)f'(u)=-\Gamma_{u\theta\theta}=-\Gamma_{\theta u\theta},\nonumber\\
\Gamma_{\phi\phi u}&=&-f(u)f'(u) \sin^2 \theta=-\Gamma_{u\phi\phi}=-\Gamma_{\phi u\phi},\nonumber\\
\Gamma_{\phi\phi\theta}&=&-
f^2(u)\sin\theta\cos\theta=-\Gamma_{\theta\phi\phi}=-\Gamma_{\phi \theta\phi}\nonumber\\
u\leq  u_0&:&\nonumber\\
\hat{\Gamma}_{uuu}&=&- F(u) F'(u),\nonumber\\
\hat{\Gamma}_{\theta\theta u}&=&- H(u)H'(u)=-\hat{\Gamma}_{\theta u\theta}=-\hat{\Gamma}_{u\theta\theta},\nonumber\\
\hat{\Gamma}_{\phi\phi u}&=&-H(u)H'(u) \sin^2 \theta=-\hat{\Gamma}_{\phi u\phi}=-\hat{\Gamma}_{u\phi\phi},\nonumber\\
\hat{\Gamma}_{\phi\phi\theta}&=&-H^2(u)\sin\theta\cos\theta=-\hat{\Gamma}_{\phi\theta\phi}=-\hat{\Gamma}_{\theta\phi\phi}
\end{eqnarray}
 We shall exploit the symmetry of the metric (\ref{metric3}) to choose the equatorial slice $\theta=\frac{\pi}{2}$ as the plane of motion. Using the expressions for the metric and affine connection given above, the geodesic equations at $u>u_0$ become:
\begin{eqnarray}
&&\left(1-\frac{2M}{f(u)}\right)\dot{t}^2-\frac{f'^2(u)\dot{u}^2}{\left(1-\frac{2M}{f(u)}\right)}-f^2(u)\dot{\phi}^2=k,\nonumber\\
&& \left(1-\frac{2M}{f(u)}\right)\dot{t}=E,~f^2(u)\dot{\phi}=L,
\end{eqnarray}
$E,L$ being the constants of motion. 
This implies (ignoring the integration constant):
\begin{eqnarray}
\lambda = \pm \int du~\left[\frac{f^3(u)f'^2(u)}{E^2 f^3(u)-[L^2+k f^2(u)][f(u)-2M]}\right]^\frac{1}{2} 
\end{eqnarray}
In the degenerate region, on the other hand, the geodesic equations along the dynamical (non-null) coordinates read:
\begin{eqnarray}
F^2(u)\dot{u}^2-H^2(u) \dot{\phi}^2 =k,~H^2(u)\dot{\phi}=L, 
\end{eqnarray}
leading to the solution:
\begin{eqnarray}
\lambda &=& \pm \int du~\frac{F(u)H(u)}{\left(L^2+k H^2(u)\right)^\frac{1}{2} }
\end{eqnarray}

Instead of working with these general solutions, we shall  consider the timelike geodesics ($k=1$) for the simpler case of $L=0$ which would be sufficient for our purpose here. The corresponding solution for the affine parameter is given by:
\begin{eqnarray}
\lambda &=& \pm \int du~\left[\frac{f(u)}{(E^2-1)f(u)+2M}\right]^\frac{1}{2}f'(u)\nonumber\\
& =& \pm\frac{f^{\frac{1}{2}}(u) [(E^2-1)f(u)+2M]^{\frac{1}{2}}}{E^2-1}\nonumber\\
&\mp &\frac{2M}{(E^2-1)^{\frac{3}{2}}} ln
[(E^2-1)^{\frac{1}{2}}\left((E^2-1)f(u)+2M\right)^{\frac{1}{2}}\nonumber\\&+&(E^2-1)f^{\frac{1}{2}}(u)]\mathrm{~~(at}~~u>u_0),\nonumber\\
&=&\pm \left[ f^{\frac{1}{2}}(u) [2M-f(u)]^{\frac{1}{2}}-2M \tan^{-1}\left[\frac{f^{\frac{1}{2}}(u)}{[2M-f(u)]^\frac{1}{2}}\right]\right]\nonumber\\
&&\mathrm{(at}~~u\leq u_0)
\end{eqnarray}
We may now fix the function $f(u)$, any legitimate choice (consistent with the boundary conditions (\ref{fF2})) of which represents an acceptable solution. Let us choose:
\begin{eqnarray}
f(u)=2M \left[1+\left(\frac{u}{u_0}-1\right)^ne^{\left[n\left(\frac{u}{u_0}-1\right)\right]}\right], 
\end{eqnarray}
where $n\geq 3$ is an odd integer. With this, the affine parameter exhibits a turning point at $u=0$ where $f(u)$ has a minimum. If an observer is able to cross this point along the geodesic, the proper clock would start running backwards. 
%It is worth mentioning that the torsional generalizations of these extended Schwarzschild geometries \cite{kaul2} are also expected to exhibit the same causal features.

Since both the tetrad and connection fields are independent of the null coordinate, the geometry of the degenerate region may be completely characterized in terms of the nondegenerate three-metric, as in the earlier examples. The associated three-curvature scalars are finite there. For instance,
\begin{eqnarray*}
 &&\bar{R}=0,~
 \bar{R}_{ab}^{~~ij} \bar{R}^{ab}_{~~ij}=\frac{8}{F^2(u) H^2(u)}\left[\frac{H'(u)}{F(u)}\right]'^2\nonumber\\
 &&+\frac{4}{H^4(u)}\left[1+\frac{H'^2(u)}{F^2(u)}\right]^2=
 \frac{8M}{f^5(u)}\left[1+\frac{M}{f(u)}\right]
\end{eqnarray*}

\section{Final remarks}

In this article, we have elucidated a general method to construct spacetime solutions in first order gravity, which admit the possibility of time travel through their geodesics. These four-geometries exhibit invertible as well as noninvertible tetrad fields over different spacetime regions, corresponding to the two possible phases of first order formulation, respectively. At the region with the noninvertible phase, there exist turning points of the affine parameter for timelike and null geodesics. Thus, a proper clock (along a timelike geodesic) which might cross any of these points must start running backwards in time.
%At the region with the noninvertible phase, an observer falling through the geodesic can revisit their past according to the clocks they carry along. 

The spacetime geometries discussed here are geodesically incomplete. Each of them contains at least one isolated point where the four-geometry is not time-orientable. On the other hand, these solutions are free of any divergence in the curvature two-form fields or in the lower dimensional curvature scalars that could be associated with the zero-determinant phase. Also, these satisfy the energy conditions \cite{samuel} .
% or associated three-curvature scalars that could be defined for the noninvertible phase. 
%Hence, the tidal accelerations between neighbouring geodesics, as encoded by the field strength, are also expected to be finite everywhere. 

It is important to emphasize that one must be cautious in demanding that the acausal solutions presented here could be actual `time machines', which we certainly would not do at this stage.  In other words, the question that whether an observer can go across the turning points of the affine parameter or the phase boundary between the degenerate and nondegenerate phase in reality remains open, in view of some of the issues mentioned above.  From the analysis of the geodesics here, the  torsionless solutions seem to exhibit causality violation only if they are not time-orientable at least at one point. This may or may not be the case for torsionful solutions of first order gravity in vacuum.

The solutions here have been constructed directly within the Hilbert-Palatini Lagrangian framework, where the covariant metric is allowed to be degenerate over a region. In general, these are not necessarily the same in essence as the solutions that have been obtained earlier within Ashtekar's Hamiltonian formulation \cite{bengtsson,bengtsson1, madhavan}. In fact, some of these geometries, where the canonical triad field exhibits a degeneracy, are known to be associated with negative energy \cite{madhavan,samuel}. It would certainly be worthwhile to explore if acausal spacetimes similar to the ones here could be realized in such a canonical framework as well.

In view of the classical solutions to the field equations presented here, it seems reasonable not to rule out the possibility of having to accomodate causality violation even within classical gravity. Such spacetimes, allowing a continuous transition to zero signature, should also be  relevant in a formulation of quantum gravity  where change of signature or of topology \cite{horowitz}  could play an important role.

%\acknowledgments
\acknowledgments 

The author is supported by (in part) grant no. ECR/2016/000027 (ECR award), SERB, Department of Science and Technology, Government of India and (in part) the ISIRD grant (code-RAQ). I thank Joseph Samuel for many useful conversations on this work and Sayan Kar for discussions on Misner space.

\end{document}